\documentclass[5p,twocolumn,times,number]{elsarticle}

\usepackage{graphicx}
\usepackage{amsmath}

\usepackage{lineno}
\linenumbers

\begin{document}

\begin{frontmatter}

\title{Planar Pixel Sensors for the ATLAS tracker upgrade at HL-LHC}

\author[add1]{C.~Gallrapp\corref{cor}}
\ead{christian.gallrapp@cern.ch}
\author{on behalf of the ATLAS Planar Pixel Sensor R\&D Project}

\cortext[cor]{Corresponding author}

\address[add1]{CERN, Geneva 23, CH-1211, Switzerland}

\begin{abstract}
The ATLAS Planar Pixel Sensor R\&D Project is a collaboration of 17 institutes and more than 80 scientists. Their goal is to explore the operation of planar pixel sensors for the tracker upgrade at the High Luminosity-Large Hadron Collider (HL-LHC).\\
This work will give a summary of the achievements on radiation studies with n-in-n and n-in-p pixel sensors, bump-bonded to ATLAS FE-I3 and FE-I4 readout chips. The summary includes results from tests with radioactive sources and tracking efficiencies extracted from test beam measurements. Analysis results of ${2\cdot10^{16}}~\text{n}_{\text{eq}}\text{cm}^{-2}$ and ${1\cdot10^{16}}~\text{n}_{\text{eq}}\text{cm}^{-2}$ ($1~\text{MeV}$ neutron equivalent) irradiated n-in-n and n-in-p modules confirm the operation of planar pixel sensors for future applications.
\end{abstract}

%
%

\end{frontmatter}

\section{Introduction}

The current ATLAS Pixel Detector was built to enable track reconstruction up to a LHC peak luminosity of 
${10^{34}}~\text{cm}^{-2}\text{s}^{-1}$~\cite{Aad:2008}. Ongoing upgrade efforts of the LHC will result in an increased peak luminosity of ${10^{35}}~\text{cm}^{-2}\text{s}^{-1}$. This will make it necessary to enhance the existing sensor technologies in terms of radiation hardness and occupancy.\\
The current ATLAS Pixel Detector was developed on a planar silicon technology. To benefit from this knowledge the \mbox{ATLAS} Planar Pixel Sensor (PPS) R\&D Project was founded to advance the research on planar pixel sensor upgrade. This included the developments for the Insertable B-Layer (IBL)~\cite{IBL:2010} but also the enhancements for the HL-LHC upgrade. Five research topics were defined by the member institutes to address this challenge from different angles~\cite{PPS:2008}:
\begin{enumerate}
	\item Choice of bulk material and radiation damage related studies
	\item Development of low-cost planar silicon pixel detectors
	\item Reduction of inactive sensor area (active edges, slim edges)
	\item Simulation to optimize the pixel cells properties 
	\item Development of analog read-out chips for low threshold operation
\end{enumerate}
Different pixel concepts for inner and outer layers of future tracking detectors are currently under investigation.\\
Extreme radiation hardness is necessary for the application of detectors in the inner layers of an experiment. This can be achieved by the application of a very thin sensor material which reduces the effects of the radiation damage. Slim and active edge designs can be used to provide low geometrical inefficiency at the module edges.\\
Cost reduction is the research focus for pixel sensors in the outer layers of the pixel detector. Production on $6^{\prime\prime}$ instead of $4^{\prime\prime}$ wafers and more cost-efficient and industrialized interconnection techniques are two of these approaches. The upcoming n-in-p technology is a promising candidate which reduces production cost by using a single sided production process.
\section{Radiation damage studies on different bulk materials}
Currently there are two planar pixel technologies under investigation. N-in-n is the technology currently used in the \mbox{ATLAS} Pixel Detector and based on a n-bulk silicon whereas the n-in-p technology uses p-bulk material. N$^+$ doping is used for the pixel implants on one side of the sensor whereas a p$^+$ doping builds the bias voltage contact on the other side.\\
Planar pixel samples from both technologies were irradiated with protons and neutrons and tested on the ATLAS FE-I3~\cite{Peric:2006} and FE-I4~\cite{Garcia-Sciveres:2011} read-out chips. Expected fluences for IBL and HL-LHC are ${5\cdot10^{15}}~\text{n}_{\text{eq}}\text{cm}^{-2}$ and ${2\cdot10^{16}}~\text{n}_{\text{eq}}\text{cm}^{-2}$ respectively. These fluences were reached by irradiation campaigns in Ljubljana~\cite{LU:irrad}, Karlsruhe~\cite{KA:irrad} and CERN~\cite{PS:irrad}.
\subsection{N-in-n pixel sensor}
\begin{figure}[h]
	\centering
	\includegraphics[width=0.99\linewidth]{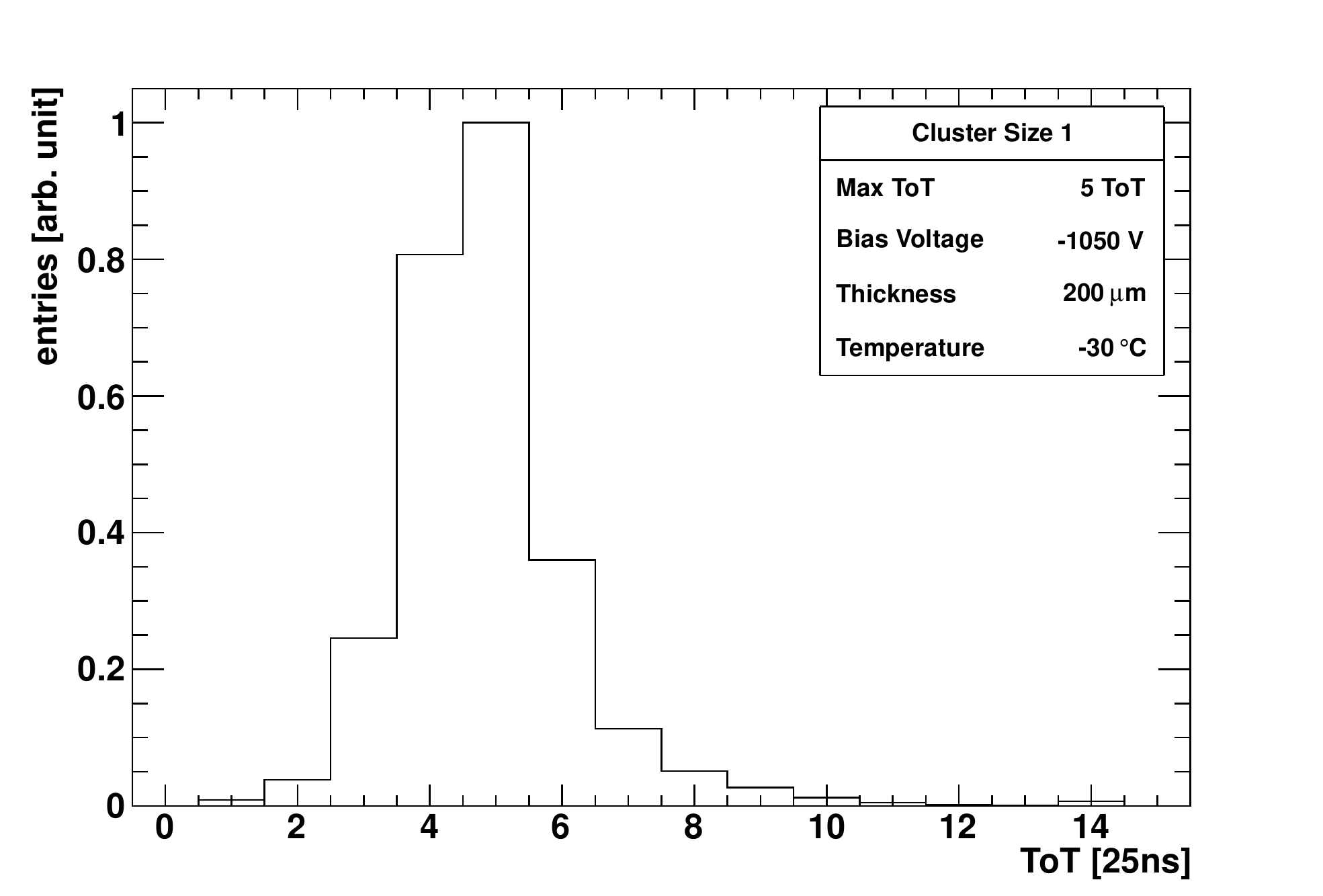}
	\caption{Charge collection for single pixel cluster of a ${5\cdot10^{15}}~\text{n}_{\text{eq}}\text{cm}^{-2}$	proton irradiated module (PS1) with a threshold tuning of 	$1.6~\text{ke}^-$}
	\label{fig:Do}
\end{figure}

Planar pixel sensor for the IBL project were tested in different steps up to the IBL requirements. The slim edge design and the reduced number of guard rings made it possible to decrease the inactive area to $200~\mu{}\text{m}$~\cite{Altenheiner:2012}. Edge efficiency studies on $250~\mu{}\text{m}$ thick and ${4\cdot10^{15}}~\text{n}_{\text{eq}}\text{cm}^{-2}$ neutron irradiated samples showed an efficiency drop very close to the pixel edge at $800~\text{V}$. Samples with a thickness of $200~\mu{}\text{m}$ are expected to show a better edge efficiency caused by the increased depletion zone at the same bias voltage.\\
The charge collection of a Sr-90 source scan with an IBL fluence proton irradiated sample is shown in Fig.~\ref{fig:Do}. The peak of the Landau distribution at a Time-over-Threshold (ToT) of $5~\text{ToT}$ is equivalent to a charge of about $8.3~\text{ke}^-$. This exceeds the threshold of $1.6~\text{ke}^-$ by more than $6.5~\text{ke}^-$ and shows that the sensors can be operated at IBL fluences.\\
\\
For HL-LHC irradiated samples a charge collection of $4~\text{ke}^-$ has been demonstrated during beam measurements at a bias voltage of $1000~\text{V}$~\cite{Altenheiner:2011,Weingarten:2012}.
\subsection{N-in-p pixel sensor}
Two types of n-in-p pixel sensors produced in Japan and Germany have been irradiated and tested with radioactive sources and beam measurements.\\

Different bias structures in combination with two isolation structures have been tested on the $150~\mu{}\text{m}$ thick sensors produced in Japan. A punch-through dot and a polysilicon resistor are the two tested bias structures. Common and individual p-stop were used as isolation structures between the pixel implants~\cite{Unno:2012}. First irradiation tests have been performed on two samples after a proton irradiation at a fluence of ${2\cdot10^{15}}~\text{n}_{\text{eq}}\text{cm}^{-2}$.\\
\begin{figure}[h]
	\centering
	\includegraphics[width=0.9\linewidth]{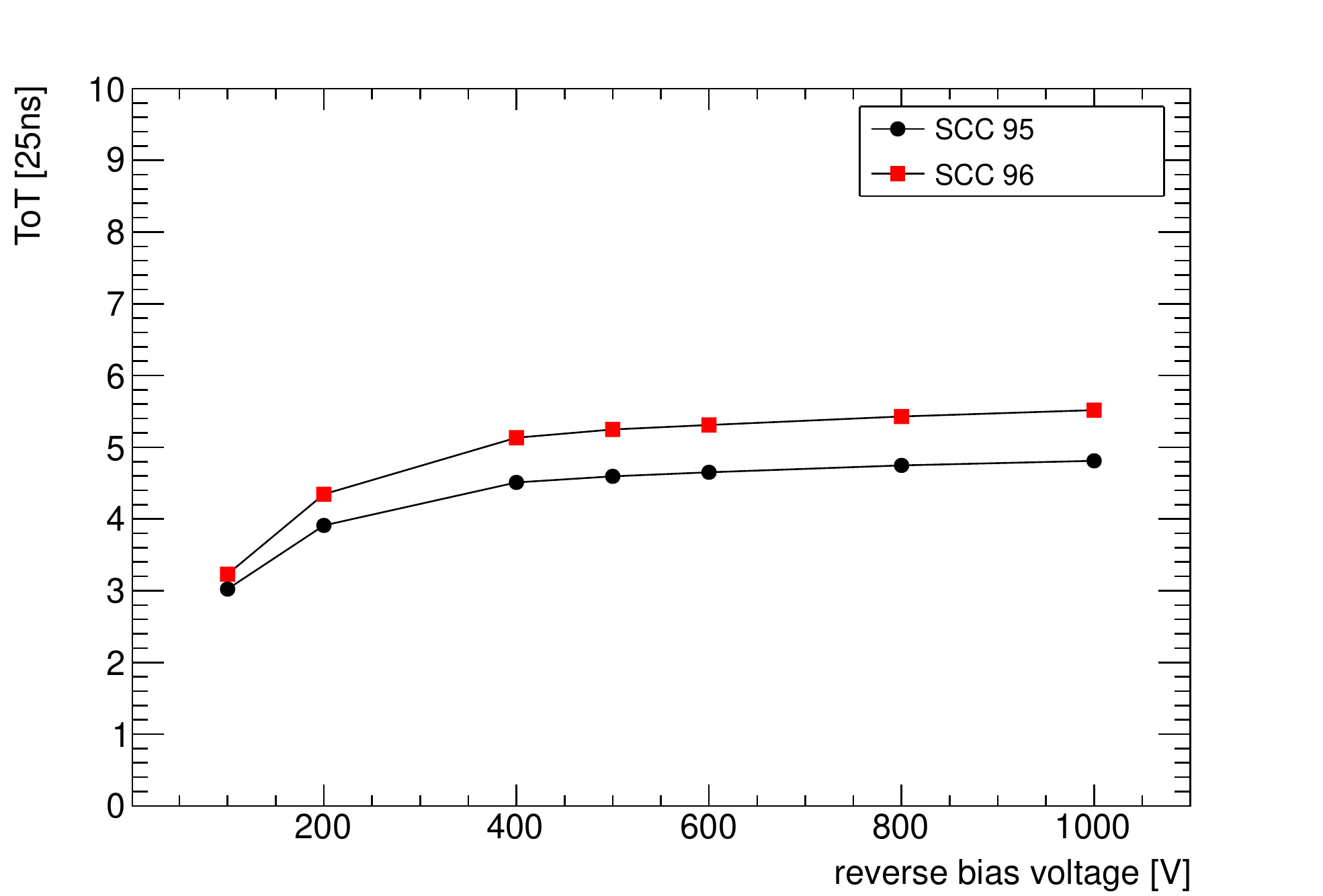}
	\caption{Most probable charge collection for single pixel cluster of two ${2\cdot10^{15}}~\text{n}_{\text{eq}}\text{cm}^{-2}$ irradiated modules as a function of the bias voltage with a threshold tuning of $1.6~\text{ke}^-$. SCC 95 is fabricated with a polysilicon resistor and common p-stop while SCC 96 uses punch-through dots and individual \mbox{p-stop}\cite{Nagai:2012}.}
	\label{fig:KEK}
\end{figure}\\
An overall efficiency at about $95~\%$ was observed for both irradiated samples at a bias voltage of $1000~\text{V}$. Although both samples showed regions with low efficiency at the pixel corners the global efficency already exceeded $90~\%$ at $200~\text{V}$. The most probable charge of more than $8~\text{ke}^-$ resulting from a Sr-90 source scan at a bias voltage of $1000~\text{V}$ for the irradiated samples is shown in Fig.~\ref{fig:KEK}. The threshold excess of more than $6.5~\text{ke}^-$ shows that these samples are promising candidates for higher fluences~\cite{Nagai:2012}.\\

Two different guard ring-designs have been fabricated in Germany on $300~\mu{}\text{m}$ thick sensors and irradiated with protons and neutrons.\\
A charge collection of more than $6.4~\text{ke}^-$ was presented for \mbox{Sr-90} source measurements on a ${1\cdot10^{16}}~\text{n}_{\text{eq}}\text{cm}^{-2}$ irradiated sample at bias voltages higher than $950~\text{V}$~\cite{LaRosa:2012}. An overall detection efficiency of $98.6~\%$ was measured during a beam test on a ${5\cdot10^{15}}~\text{n}_{\text{eq}}\text{cm}^{-2}$ irradiated sample at a bias voltage of $600~\text{V}$~\cite{Gallrapp:2012}.
\section{Summary and Outlook}
The PPS collaboration showed that planar pixel sensors can be operated at intermediate fluences, as for the IBL, but also at HL-LHC fluences. Designs with reduced inactive area have been successfully tested and are still improving. Further irradiation campaigns are already planed to gain more statistics about the performance at high fluences.
\section*{Acknowledgments}
The author would like to thank A. Dierlamm (KIT), I. Mandic (JSI) and M. Glaser (CERN) for carrying out the device irradiation in Karlsruhe, Ljubljana and CERN respectively. The irradiation in Karlsruhe was supported by the Initiative and Networking Fund of the Helmholtz Association, contract HA-101. We would like to thank all the members of the PPS testbeam crew for their support during the testbeam campaigns of the last years. The research leading to these results has received funding from the European Commission under the FP7 Research Infrastructures project AIDA, grant agreement no. 262025.

\end{document}